\documentclass[12pt]{article}

\usepackage{graphicx}
\usepackage{bm}
\usepackage{amsmath}
\usepackage{amssymb}
\usepackage[bbgreekl]{mathbbol}

\def\be{\begin{equation}}
\def\ee{\end{equation}}
\def\bea{\begin{eqnarray}}
\def\eea{\end{eqnarray}}

\begin{document}

\pagestyle{plain}

\begin{center}
~

\vspace{2cm} {\large \textbf{Coordinate/Field Duality in Gauge Theories: Emergence of Matrix Coordinates}}

\vspace{1cm}

Amir H. Fatollahi

\vspace{.5cm}

{\it Department of Physics, Alzahra University, \\ P. O. Box 19938, Tehran 91167, Iran}

\vspace{.3cm}

\texttt{fath@alzahra.ac.ir}

\vskip .5 cm
\end{center}

\begin{abstract}
The proposed coordinate/field duality [Phys. Rev. Lett. \textbf{78} (1997) 163]
is applied to the gauge and matter sectors of gauge theories. In the non-Abelian
case, due to indices originated from the internal space, the dual coordinates
appear to be matrices. The dimensions and the transformations of
the matrix coordinates of gauge and matter sectors are different and are
consistent to expectations from lattice gauge theory and the theory of
open strings equipped with the Chan-Paton factors.
It is argued that in the unbroken symmetry phase,
where only proper collections of field components as colorless states are
detected, it is logical to assume that the same happens for the dual
coordinates, making matrix coordinates the natural candidates to capture
the internal dynamics of baryonic confined states. The proposed matrix coordinates
happen to be the same appearing in the bound-state of D0-branes of string theory.
\end{abstract}

\vspace{1cm}


\newpage
According to one interpretation of the special relativity agenda, it would be
meaningful to demand for similar characters between the coordinates of
space-time and the fields living on it. In particular, as far as the propagation of
electromagnetic waves is concerned, one may assume that the space-time
coordinates as well as the electromagnetic potentials should transform similarly;
then, as an immediate consequence, it is expected that under boost transformations
the space and time coordinates would be mixed. Also by this way of interpretation,
the super-space formulation of supersymmetric theories is just a natural continuation
of the special relativity program: The inclusion of anti-commutating coordinates as
representatives of the fermionic degrees of freedom of the theory.

In \cite{xpsi} a duality between coordinates of space-time and fields is
formulated based on the inversion of the field equation. In particular the
prepotential $\mathcal{F}^{(\mu)}$ relating two linear independent
solutions would appear as the Legendre transform of the coordinate $x^\mu$ with respect
to the field density \cite{xpsi}. It is argued that this duality may generate new structures
upon second quantization of fields, leading to the quantized version of space-time coordinates.

In \cite{vancea} the above mentioned duality was applied to the spinor fields.
In particular it is observed that in the case of Dirac field, for which the field
has extra indices of $spin(1,3)$, the proposed duality can induce the same extra
structure on the dual space-time coordinates as well, evolving them into
matrices \cite{vancea}. It is remarked that upon second quantization, where
the spinor components would become anti-commuting, the same character can
also be induced on the coordinates as well \cite{vancea}.

In the present work the proposed duality is examined in case of gauge theories.
In particular, the consequences of the extra structures coming
from the internal space and the transformations associated to them on the
coordinates dual to the fields are investigated.

Let us begin with the Abelian case, in which the U(1) field equations read
\begin{align}\label{1}
\partial^\mu F_{\mu\nu}=j_\nu
\end{align}
where $F_{\mu\nu}=\partial_\mu A_\nu-\partial_\nu A_\mu$, and
$j_\mu$ is the source coming from the matter sector of the theory.
For example in a theory with Dirac spinors as matter content
$j_\mu=g\, \bar{\psi}\gamma_\mu\psi$ with $g$ as coupling constant.
The theory is symmetric under the local gauge transformations:
\begin{align}\label{2}
A'_\mu=A_\mu+\frac{1}{g}\partial_\mu \chi,~~~~~~\psi'=e^{i \chi}\psi
\end{align}
leading to $j'_\mu=j_\mu$, where $\chi$ is an arbitrary differentiable function.
The physical degrees of freedom are obtained once one uses the gauge symmetry to
eliminate the redundant degrees, for which we adopt the Lorentz gauge
$\partial\cdot A=0$, together with the condition that the time component of
the gauge field vanishes. For the free field ($j_\mu=0$) this leads to $\square A_\mu=0$ with
solutions
\begin{align}\label{3}
\bm{A}_{(\lambda)}=\bm{\epsilon}_{(\lambda)}\,e^{-i\, k\cdot x},~~~ A_0=0
\end{align}
in which $k\cdot x=k^0 x^0 - \bm{k}\cdot\bm{x}$, and
$\bm{\epsilon}_{(\lambda)}$ represents two right and left circular
polarization vectors ($\lambda=R~ \&~ L$) satisfying the transversity condition
$\bm{\epsilon}_{(\lambda)}\cdot \bm{k}=0$. The linearly independent solution,
$\widetilde{A}$, is simply obtained by the complex conjugation, that is
$\widetilde{A}=A^\ast$. The construction of the prepotential
can be done in the same line described for the Klein-Gordon field \cite{xpsi},
except that here for each polarization the construction can be done:
\begin{align}\label{4}
\widetilde{\bm{A}}_{(\lambda)}^{(\mu)}=\frac{\partial\mathcal{F}_{(\lambda)}^{(\mu)}}
{\partial \bm{A}_{(\lambda)}^{(\mu)}},~~~~\lambda=R~\&~L
\end{align}
in which the space-time index $\mu$ is emphasizing that the above is
specialized for the space-time coordinate $x^\mu$ \cite{xpsi},
that is
\begin{align}\label{5}
x^\mu_{(\lambda)}\leftarrow
\mathop{~}^\mathrm{Legendre}_\mathrm{Transform}
\rightarrow \mathcal{F}_{(\lambda)}^{(\mu)}
\end{align}
Now the important point is that by the two gauge fixing conditions
($\partial\cdot A=0$ and $A_0=0$), apart from functions which diverge
at infinity or are constant, there is no further choice to change
the gauge fields by (\ref{2}). In other words, as the gauge symmetry is almost
totally fixed, there is no need to examine the effect of the gauge transformations
on the space-time coordinates dual to the physical degrees of freedom
by fields.

The above situation with U(1) theory is dramatically changed for the non-Ableian case, for which
the field equations are
\begin{align}\label{6}
\partial^\mu \mathbf{F}_{\mu\nu} -i\, g\,[\mathbf{A}^{\mu},\mathbf{F}_{\mu\nu}]=\mathbf{j}_\nu
\end{align}
where $\mathbf{F}_{\mu\nu}=\partial_\mu \mathbf{A}_\nu-\partial_\nu \mathbf{A}_\mu
-i\, g\, [\mathbf{A}_{\mu},\mathbf{A}_{\nu}]$, the gauge potential
$\mathbf{A}_\mu$ and the source $\mathbf{j}_\mu$ are matrix-valued quantities in
the group algebra. For example the gauge potential and the field strength have
the usual expansion in group algebra
\begin{align}\label{7}
\mathbf{A}_\mu=A_\mu^a\,T_a,~~~~\mathbf{F}_{\mu\nu}=F_{\mu\nu}^a\,T_a
\end{align}
in which the group generators $T_{a}$'s satisfy the commutation relation
\begin{align}\label{8}
\big[ T_{a},T_{b}\big]= i\, f^c_{~ab}\,T_{c}
\end{align}
Again the theory is invariant under the gauge transformations
\begin{align}\label{9}
\mathbf{A}'_\mu=U \mathbf{A}_\mu U^\dagger+\frac{i}{g}U\partial_\mu U^\dagger,~~~~~~\bm{\psi}'=U \bm{\psi }
\end{align}
where $U=\exp(i\, \bm{\chi} )$ with $\bm{\chi}=\chi^a\,T_a$.
One may use the the gauge fixing conditions
\begin{align}\label{10}
\partial\cdot A^a=0, ~~ A_0^a=0,~~~~~ \forall~a
\end{align}
by which the solutions of the field equations would represent the
physical degrees of freedom. For example, in the zero coupling limit
$g\to 0$, one has $\square \mathbf{A}_\mu\approx 0$ and $\square \bm{\psi}\approx 0$,
with plane-waves as solutions. From now on to treat indexing of the
gauge and matter fields' components similarly, we adopt the notation by which
the elements of matrices representing the fields happen to appear in expressions.
It is convenient to define the set $\{\xi_\alpha\}$ of column basis vectors by components:
$(\xi_\alpha)^\beta=\delta_\alpha^\beta$. In particular,
for the group U($N$) in the defining representation, for the $N\times N$ dimensional matrix
$\mathbf{A}$ and the $N$-dimensional vector $\bm{\psi}$, we have
\begin{align}\label{11}
\mathbf{A}&=\sum_{\alpha,\beta} A_{\alpha\beta}~
\xi_\alpha\otimes \xi_\beta^T\\
\label{12}
\bm{\psi}&=\sum_\alpha \psi_\alpha~\xi_\alpha
\end{align}
with superscript $T$ for transpose operation, and
$\alpha,\beta=1,\cdots, N$ (for the index in (\ref{7}) we have $a=1,\cdots,N^2$).
By this for the zero coupling limit in an obvious matrix notation
we may represent the solutions as:
\begin{align}\label{13}
\mathbf{A}_{\alpha\beta}&\propto{\xi}_\alpha\otimes{\xi}_\beta^T~e^{-i\, k\cdot x}\\
\label{14}
\bm{\psi}_\alpha&\propto\xi_\alpha~e^{-i\, k\cdot x}
\end{align}
in which for the sake of brevity we have ignored the information about
the polarizations of gauge and matter fields.
For the finite but small gauge coupling one may use the perturbative
techniques to solve the coupled differential equations of fields.
In this setup the linearly independent solutions are obtained simply by
$\dagger$-operation, recalling $\xi^\dagger=\xi^T$, in the zero coupling limit we have
\begin{align}\label{15}
\widetilde{\mathbf{A}}_{\alpha\beta}&\propto{\xi}_\beta\otimes{\xi}_\alpha^T~e^{i\, k\cdot x}\\
\label{16}
\widetilde{\bm{\psi}}_\alpha&\propto\xi_\alpha^T~e^{i\, k\cdot x}
\end{align}
Following \cite{xpsi,vancea} by the two linearly independent solutions one may define
the prepotentials, by which the duality between the matrix/vector
fields and the coordinates can be constructed:
\begin{align}\label{17}
\widetilde{\mathbf{A}}_{\alpha\beta}^{(\mu)}&=\frac{\partial\mathcal{F}_{\alpha\beta}^{(\mu)}}
{\partial \mathbf{A}_{\alpha\beta}^{(\mu)}},\\
\label{18}
\widetilde{\bm{\psi}}^{(\mu)}_\alpha&=\frac{\partial\mathcal{G}^{(\mu)}_\alpha}
{\partial \bm{\psi}^{(\mu)}_\alpha},
\end{align}
again index $\mu$ is indicating that the above is for the dual coordinate
$x^\mu$ \cite{xpsi}. In order to express the dual coordinates as functions of
fields and the prepotentials it is needed to differentiate 
from prepotentials with respect to $x^\mu$, following \cite{vancea} we have
\begin{align}\label{19}
\partial_\mu \mathcal{F}_{\alpha\beta}^{(\mu)}&=
\frac{\partial \mathbf{A}_{\alpha\beta}^{(\mu)}}{\partial x^\mu}\otimes
\frac{\partial\mathcal{F}_{\alpha\beta}^{(\mu)}}
{\partial \mathbf{A}_{\alpha\beta}^{(\mu)}}
\\
\label{20}
\partial_\mu \mathcal{G}_{\alpha}^{(\mu)}&=
\frac{\partial \bm{\psi}_{\alpha}^{(\mu)}}{\partial x^\mu}\otimes
\frac{\partial\mathcal{G}_{\alpha}^{(\mu)}}{\partial \bm{\psi}_{\alpha}^{(\mu)}}
\end{align}
by which the dual coordinates are defined through the Legendre transform \cite{vancea}
\begin{align}\label{21}
x^{\mu}_{\alpha\beta,\alpha\beta}
&=\frac{1}{2}\mathbf{A}_{\alpha\beta}\otimes \frac{\partial\mathcal{F}_{\alpha\beta}^{(\mu)}}
{\partial \mathbf{A}_{\alpha\beta}^{(\mu)}}- \mathcal{F}_{\alpha\beta}^{(\mu)}+C^{(\mu)}_{\alpha\beta}\\
\label{22}
x^{\mu}_{\alpha\alpha}
&=\frac{1}{2}\bm{\psi}_{\alpha}^{(\mu)}\otimes \frac{\partial\mathcal{G}^{(\mu)}_\alpha}
{\partial \bm{\psi}^{(\mu)}_\alpha} - \mathcal{G}^{(\mu)}_\alpha +D^{(\mu)}_{\alpha}
\end{align}
with definitions \cite{vancea}
\begin{align}\label{23}
x^{\mu}_{\alpha\beta,\alpha\beta}&=(\xi_\alpha\otimes\xi_\beta^T)\otimes
(\xi_\beta\otimes\xi_\alpha^T)~x^\mu\\
\label{24}
x^{\mu}_{\alpha\alpha}&=(\xi_\alpha\otimes\xi_\alpha^T)~x^\mu
\end{align}
Some comments are in order. First, the numbers of coordinates of gauge and matter sectors
are equal to the number of the fields' components in each sector;
$N^2$ for the gauge fields and $N$ for the matter fields.
Second, the matrix coordinates for the gauge and matter sectors are
matrices of $N^2\times N^2$ and $N\times N$ dimensions, respectively,
with the only non-zero elements as
\begin{align}\label{25}
x^{\mu}_{\alpha\beta,\alpha\beta}&:((\alpha-1)N+\beta,(\beta-1)N+\alpha)\\
\label{26}
x^{\mu}_{\alpha\alpha}&:(\alpha,\alpha)
\end{align}
It is interesting to note that the matrix coordinates dual to the matter fields are diagonal.
Third, in contrast to the case with column spinors in \cite{vancea},
here the matrices arising from the tensor products are not invertible. As the consequence,
one can not get rid of the matrix structure by simply multiplying the two sides
of the Legendre transform by the inverse matrix \cite{vancea}.

Now one may consider the effect of the gauge transformations of fields on
the dual coordinates. In contrast to the case with Abelian symmetry, here the
global gauge transformations are not trivial, causing the mixing between
the different components of fields in internal space, explicitly
for the fields' components by (\ref{9})
\begin{align}\label{27}
{A}'_{\alpha\beta}&=\sum_{\gamma,\delta=1}^N U_{\alpha\gamma}\,A_{\gamma\delta}\,U^\dagger_{\delta\beta},\\
\label{28}
\psi'_\alpha&=\sum_{\beta=1}^N U_{\alpha\beta}\,\psi_\beta
\end{align}
and $\dagger$-operated of above for the linearly independent
solutions $\widetilde{A}$ and $\widetilde{\psi}$. Referring to
(\ref{21}) and (\ref{22}) we find the transformation rules for the dual coordinates as
\begin{align}\label{29}
x'^{\mu}_{\alpha\beta,\alpha\beta}&=(U\xi_\alpha\otimes\xi_\beta^T U^\dagger)\otimes
(U\xi_\beta\otimes\xi_\alpha^TU^\dagger)~x^\mu\\
\label{30}
x'^{\mu}_{\alpha\alpha}&=(U\xi_\alpha\otimes\xi_\alpha^TU^\dagger)~x^\mu
\end{align}
or in a matrix notation
\begin{align}\label{31}
{x}'^\mu_{\alpha\beta,\alpha\beta}&=U\otimes U \,
{x}^\mu_{\alpha\beta,\alpha\beta}\, U^\dagger \otimes U^\dagger\\
\label{32}
{x}'^\mu_{\alpha\alpha}&=U\,{x}^\mu_{\alpha\alpha}\, U^{\dagger}
\end{align}
In above one notices that the coordinates dual to the gauge fields transform as
if they were consisting double copies of the matter labels. This observation is
consistent with the fact that in lattice formulation of gauge theories, while the
matter degrees of freedom are associated to the sites, the gauge degrees of
freedom are defined on the links joining the sites \cite{lattice}. In particular, the
matter sectors sitting on two adjacent sites $I$ and $J$ share their labels with the
gauge degrees of freedom living on the link connecting these two sites.
This picture is also fully consistent with the way that Chan-Paton factors are
introduced in the open string theory. In particular the matter (quark) labels attached to two
ends of an open string would make it acting as a gauge field degrees of freedom
\cite{chanpaton}.

There is the possibility by which one can represent the coordinates dual to the gauge fields
more economically. In fact each pair $x^{\mu}_{\alpha\beta,\alpha\beta}$ and
$x^{\mu}_{\beta\alpha,\beta\alpha}$ as $N^2\times N^2$ matrices can be assembled to
$\mathbf{x}^{\mu}_{\alpha\beta}$ as $N\times N$ hermitian matrix defined by
\begin{align}\label{33}
\mathbf{x}^{\mu}_{\alpha\beta}=
\left({x}^\mu_{\alpha\beta}+({x}^{\mu}_{\alpha\beta})^T\right)
+i\left(({x}^{\mu}_{\beta\alpha})^T-{x}^{\mu}_{\beta\alpha}\right)
\end{align}
in which the reduced matrices are defined by
\begin{align}\label{34}
{x}^{\mu}_{\alpha\beta}:=\xi_\beta^T\cdot{x}^{\mu}_{\alpha\beta,\alpha\beta}\cdot\xi_\alpha
=(\xi_\alpha\otimes\xi_\beta^T)~x^\mu
\end{align}
In above one notices that $x^\mu$'s appearing in $x^{\mu}_{\alpha\beta,\alpha\beta}$ and
$x^{\mu}_{\beta\alpha,\beta\alpha}$ are in fact different,
as they are dual to different components of the
gauge field, namely $A_{\alpha\beta}$ and $A_{\beta\alpha}$.
In this way of representation, the matrix coordinates and the gauge fields
appear as $N\times N$ hermitian matrices. Further, this way of representation
seems unavoidable once one is going to represent the allowed physical states as collections of all
components of matter and gauge fields. In particular, in the unbroken symmetry phase,
in which the global gauge transformations would mix the components in each matter and gauge
sector, it is natural to assume that the proper coordinate to describe the
internal dynamics of physical states should be a collection of the coordinates dual
to entire matter and gauge sectors. We already have seen that the coordinates
dual to matter sector are real diagonal $N\times N$ matrices, hence there should be
matrices with the same size for the gauge sector to construct the collective coordinates
to describe the bound-states. By these all the hermitian matrices
\begin{align}\label{35}
\mathbf{X}^\mu:=\sum_{\alpha,\beta=1}^N \mathbf{x}_{\alpha\beta}^\mu
\end{align}
might be proposed to capture the dynamics of baryonic confined states, with the global
gauge transformation rule
\begin{align}\label{36}
\mathbf{X}'^\mu=U\, \mathbf{X}^\mu\, U^\dagger
\end{align}
The matrix coordinate constructed above is just the one which is proposed to capture the
dynamics of bound-states of D0-branes of string theory \cite{9510135}. In this picture
the diagonal elements would represent the dynamics of the D0-branes, and the
$N^2-N$ off-diagonal elements would capture the dynamics of oriented open strings
stretched between $N$ D0-branes \cite{9510135,tasi}.

In a series of works a model was considered based on the possibility
that the quantum mechanics of matrix coordinates can be used for
reproducing the features expected from
non-Abelian gauge theories. The model has shown its ability to reproduce or cover
some features and expectations in hadron physics including the potentials
between static and fast decaying quarks, the Regge behavior in
the scattering amplitudes, and possible exhibition of linear spin-mass
relation \cite{fat1,fat2,fat3}.
The symmetry aspects of the above mentioned picture were studied in \cite{fat4}.
Based on the above observations,
it is argued that maybe the matrix coordinate description of gauge theories
could generate the stringy aspects expected from gauge theories, without the need to
treat the world-sheet anomalies present in the non-critical space-time dimensions \cite{fat3,fat4}.

As the final remark, the coupling of the collective coordinates to the external potentials can be formulated as usual \cite{fat1,fat2}
\begin{align}\label{37}
S_\mathrm{int.}=q\int {\rm d}\tau \;{\rm Tr}
\left(\dot{\mathbf{X}}^\mu \mathbf{A}_\mu \right)
\end{align}
According to the D0-brane picture, the trace of matrix coordinates, which is invariant under
transformation (\ref{36}), represent the motion of the center-of-mass. By the above coupling 
we find that the center-of-mass of bound-state is not directly coupled to the traceless
parts of the potentials, just as expected for the colorless states.
However, one notices that the indirect coupling due to the inhomogeneous field strengths is possible,
just as the way that the center-of-mass of a dipole is affected by a non-constant field strength.

~\\
\textbf{Acknowledgement}:  This work is supported by
the Research Council of the Alzahra University.


\end{document}